\documentstyle[aps,tighten,12pt,prd,epsfig]{revtex}
\def \Eq#1{Eq.(\ref{#1})}
\def \Fig#1{Fig.\thinspace\ref{#1}}
\def\alem{\alpha_{\rm em}}
\def\nf{n_{\rm f}}
\def\LL{\Lambda_{\rm QCD}^2}
\def\fNS{\bar f^\gamma_{q/\bar q}(n,t,t_\gamma)}
\def\feNS{\bar f^e_{q/\bar q}(n, t, t_1, t_0)}

\begin{document}
%\preprint{TPJU-13/98}
\author{Wojciech S\l omi\'nski
%\\
\address{
Institute of Computer Science, Jagellonian University,
Reymonta 4, 30-059 Krak\'ow, Poland}}
\title{%
QCD Anomalous Structure of Electron\thanks{Work supported by
the Polish State Committee for Scientific Research
(grant No. 2~P03B~081~09) and  
the Volkswagen Foundation.}}
\maketitle

\vbox to 0pt{\vskip-80pt\hbox{TPJU-13/98}}

%-----    Abstract    ------------------------------------------------
\begin{abstract}
The parton content of the electron is analyzed within
perturbative QCD. It is shown that electron acquires an anomalous
component from QCD, analogously to photon.
The evolution equations for the `exclusive' and
`inclusive' electron structure function are constructed and solved 
numerically in the asymptotic $Q^2$ region.
\end{abstract}

%---------------------------------------------------------------------------------

\section{Introduction}

The photon structure function describes the distribution of QCD partons inside a photon. 
It is known for long \cite{resph} to have `anomalous' component, which is
calculable within perturbative QCD 
and dominates at asymptotically large momentum scales.
This asymptotic solution, as opposed to those for hadrons, is independent of 
input data measured at lower momentum scales. 
At finite scales the photon structure gets modified by
both perturbative and non-perturbative QCD contributions.

The QCD structure of the photon is revealed in interactions with
a highly virtual `probe'. 
To fix attention let us think of a virtual gluon $G^*$ with momentum $q$,
probing the photon which gets resolved into QCD partons.
Their density 
$f^\gamma_k(x,Q^2)$ ($k=q,\bar q, G$), 
depends on fractional momentum $x$ of the parton with respect to photon
and on the gluon virtuality $Q^2= |q^2|$, which must be
large as compared to the QCD
scale $\LL$.

\begin{figure}
\centerline{%
\epsfig{file=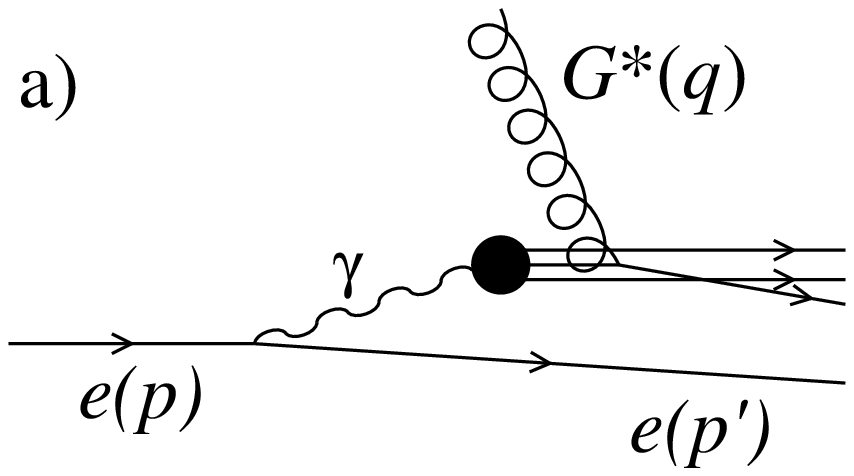,width=6.5cm}%
\hspace{10mm}
\epsfig{file=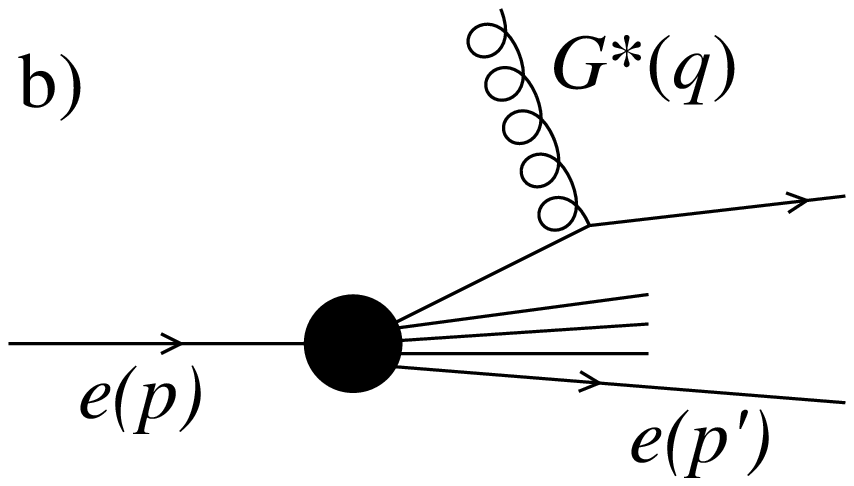,width=6.5cm}%
}
\caption{\label{F:dis}
Deep inelastic scattering on a photon (a) 
and electron (b) target}
\end{figure}

The photon structure is measured in experiments
where the electron serves as a target. 
The process is depicted in \Fig{F:dis}a, where also the notation is given.
The black blob denotes `resolved' photon and sums up all 
collinear 
QCD contributions.
The full cross-section gets also a contribution from the hard (``direct'') $G^*\gamma$
scattering (see {\it e.g.} \cite{BS}), but we will not discuss it in this paper.
Actually the photons emitted by the electron are virtual and,
from the point of view of a physical process,
$G^*$ measures the structure (parton content) of the electron, as
depicted in \Fig{F:dis}b.

%FFFFFFFFFFFFFFFFFFFFFFFFFFFFFFFFFFFFFFFFFFFFF

The aim of this paper is to study the QCD predictions for the electron
structure function at large momentum scales. 
In particular we will discuss the dependence on the 
maximal virtuality of intermediate photons emitted by the electron,
{\it i.e.} on the maximal momentum transfer between final and initial electron.
In the next section we present the problem in general and discuss the
relation to experimentally measured quantities.
In Sec. 3 we derive the QCD solution to the non-singlet electron 
structure function in the moments space. A complete set of evolution
equations is constructed and solved in Sec. 4.
In Sec. 5 we present the summary and outlook.

%SSSSSSSSSSSSSSSSSSSSSSSSSSSSSSSSSSSSSSSSSSSSSSSSSSSSSSSSSSSSSS
\section{General Framework}
The space-like virtuality of the photon exchanged in the diagram in 
\Fig{F:dis}a
\begin{equation}
  \label{gvirt}
  (p-p')^2 \equiv -P_\gamma^2
\end{equation}
can be fixed by measuring the momentum $p'$ of the outgoing electron.
Otherwise it lies within kinematic limits
\begin{equation}
  \label{Plims}
  P^2_{\rm min}(y) \equiv m_e^2 {y^2\over 1-y} 
\le P_\gamma^2 \le 
  Q^2 {z+y-zy\over z},
\end{equation}
where $m_e$ is the electron mass, $y = qp_\gamma/qp$ and 
$z = Q^2/2pq$
(these are approximate expressions
valid at $P_\gamma^2 \ll Q^2$ --- 
the exact formulae can be found in \cite{wjESF}).
Usually the upper limit on $P_\gamma^2$ is set by experimental
conditions (e.g. by anti-tagging)
\begin{equation}
  \label{uplim}
  P_\gamma^2 \le P^2.
\end{equation}
The density of partons ($k=q,\bar q, G$) with momentum $p_k = zp$ 
seen by our probe in the electron reads
\begin{mathletters}
\label{ESF0}
\begin{eqnarray}
f^e_k(z,Q^2,P^2)
&=&
  \int dx\,dy\,  \delta(z-xy)\!\!\!\!
  \int\limits_{P^2_{\rm min}(y)}^{P^2} \!\!\!\!\! dP_\gamma^2\;
  f^e_\gamma(y, P_\gamma^2)
  f^\gamma_k(x,Q^2, P_\gamma^2)
\\
&=&
  \int\limits_z^1 {dy \over y}
  \int\limits_{P^2_{\rm min}(y)}^{P^2} \!\!\!\!\! dP_\gamma^2
  f^e_\gamma(y, P_\gamma^2)
  f^\gamma_k({z \over y},Q^2, P_\gamma^2),
\end{eqnarray}
\end{mathletters}
where
\begin{equation}
  \label{dEPA}
  f^e_\gamma(y, P_\gamma^2) 
= {\alem\over 2\pi}{1\over P^2_\gamma} \left[
{1+(1-y)^2 \over y} - 2y {m_e^2\over P_\gamma^2}
\right]
\end{equation}
and
$f^\gamma_k(x,Q^2, P_\gamma^2)$ describes the $G^*\gamma $ interaction.
In \Eq{dEPA} only transverse photons are taken into account
which is 
correct within
the leading order of perturbative QCD.
All QCD contributions are contained in
$f^\gamma_k(x,Q^2, P_\gamma^2)$,
which is discussed in the literature as the structure function of
virtual photon \cite{UW,BS,GRS}.
In the following we will assume that $P^2 \gg m_e^2$ 
which allows us to neglect
the second term in the square brackets of
\Eq{dEPA}.

For
$Q^2 \gg P^2$ 
$f^\gamma_k(x,Q^2, P_\gamma^2)$  can be approximated by
the structure function of real photon ($P_\gamma^2 = 0$)
and upon integration over $P_\gamma^2$
we arrive at the
Weizs\"acker-Williams \cite{WW} formula:
\begin{equation}
  \label{W-W}
f^e_k(z,Q^2,P^2) \approx
  \int\limits_z^1 {dy \over y}
\hat f^e_\gamma(y) 
  f^\gamma_k({z \over y},Q^2, 0)
\;
  \log{P^2 \over P^2_{\rm min}(z)}
,
\end{equation}
where
\begin{equation}
\hat f^e_\gamma(y) 
=
{\alem\over 2\pi}
\, {1+(1-y)^2 \over y}.
\end{equation}
Formula (\ref{W-W}) has probabilistic interpretation in terms of
the density of photons emitted by the electron 
and the density of QCD partons within the photon.
As discussed in \cite{wjESF}, this partonic picture breaks
down at very high energies when $Z$ and $W$ bosons contribute.

In general, the experimentally measured electron structure function
is always 
integrated over a range of photon virtualities
and summed over the contributions from all weak intermediate bosons.
This structure function
describes the QCD content of a real (on-shell) electron
and allows for probabilistic interpretation
of the cross sections.
Even when we neglect the contributions from $Z$ and $W$ bosons
the integration over $P_\gamma^2$ 
disables the partonic interpretation \cite{DG}.
As compared to the standard QCD structure functions the electron one
has extra dependence on maximal photon virtuality $P^2$, which
means that we do not integrate over all final electron states.
In this sense we say that this electron structure function 
is `exclusive'.

\Eq{W-W} is an approximation to the electron structure function 
for $Q^2 \gg P^2$. 
The approach to this limit within QCD is discussed in the next section.

In the following we will assume that the $Z$ and $W$ 
bosons do not contribute but we will allow for arbitrary $P^2 \le Q^2$.
For both $Q^2$ and $P^2$ much greater than $m_e^2$
we have ({\it c.f.}{} \Eq{ESF0})
\begin{equation}
  \label{splot}
f^e_k(Q^2,P^2) =
  \int\limits_{P^2_{\rm min}}^{P^2}\! {dP_\gamma^2 \over  P_\gamma^2}
\hat f^e_\gamma \otimes
  f^\gamma_k(Q^2, P_\gamma^2),
\end{equation}
with explicit $z$ dependence suppressed and $\otimes$
denoting convolution
\begin{equation}
  \label{convdef}
  (f\otimes g)(z) \equiv
  \int\limits_0^1 dx
  \int\limits_0^1 dy
  \,\delta(z-xy)
\,f(x) g(y).
\end{equation}

We know from experiment that
a nearly real photon ($P^2_\gamma \ll \LL$) 
has a hadronic component which
is often described phenomenologically in terms
of the Vector Meson Dominance model (VDM) 
(see {\it e.g.} \cite{BS,GRS} and references therein).
This non-perturbative hadronic component
becomes less important at higher $Q^2$.
As will be shown in the next section,
any perturbative QCD predictions for $P_\gamma^2$ dependence require
$P_\gamma^2 > \Lambda_{\rm QCD}^2$ and this is the region we will
consider in details.
To this end we split the integral over $P_\gamma^2$ in \Eq{splot}
into $\int_{P_{\rm min}^2}^{P_0^2} + \int_{P_0^2}^{P^2}$ with some
$P_0^2 > \Lambda_{\rm QCD}^2$. In the first integral we use
the VDM-like photon structure function, while
the whole dependence on $P^2$ is contained in the second one:
\begin{eqnarray}
\label{qsplot}
f^e_k(Q^2,P^2) &=&
\hat f^e_\gamma \otimes
  \int\limits_{P_{\rm min}^2}^{P_0^2}\! {dP_\gamma^2 \over  P_\gamma^2}
  f^{\rm (V)}_k(Q^2, P_\gamma^2)
\nonumber\\
&&+
\hat f^e_\gamma \otimes
  \int\limits_{P_0^2}^{P^2}\! {dP_\gamma^2 \over  P_\gamma^2}
  f^\gamma_k(Q^2, P_\gamma^2).
\end{eqnarray}

%SSSSSSSSSSSSSSSSSSSSSSSSSSSSSSSSSSSSSSSSSSSSSSSSSSSSSSSS

\section{QCD calculation of electron structure function}
%===================================================

From the theoretical point of view the QCD behavior of structure functions 
is most easily analyzed in terms of their moments, defined as
\begin{equation}
f(n) = \int\limits_0^1 dx\, x^{n-1}  f(x) 
\end{equation}
for any function $f$.

The electron structure function we are going to investigate 
has the form of
the second term of \Eq{qsplot} and its moments read
\begin{equation}
\label{qsplot1}
f^e_k(n,t,t_1,t_0) =
\hat f^e_\gamma(n)
  \int\limits_{t_0}^{t_1}\! dt_\gamma
  f^\gamma_k(n,t, t_\gamma),
\end{equation}
where
\begin{equation}
t = \log{Q^2 \over \LL}\;,
\;\;
t_0 = \log{P_0^2 \over \LL}\;,
\;\;
t_1 = \log{P^2 \over \LL}\;,
\;\;
t_\gamma = \log{P_\gamma^2 \over \LL}
\end{equation}
and explicit $t_0$ argument of $f^e_k(n,t,t_1,t_0)$
is to remind on the dependence on `auxiliary' scale $P_0^2$.

Our task will be to
integrate $f^\gamma_k(n,t, t_\gamma)$ over $t_\gamma$
and to construct master equations for the electron structure function.
In order to introduce notation and 
get some understanding of the energy scales involved,
let us first briefly remind the derivation of the 
virtual photon structure function.

The master (DGLAP) equations \cite{DGLAP} read
\begin{mathletters}
\label{meg}
\begin{eqnarray}
\label{meg-q}
{df^\gamma_{q/\bar q}(n,t,t_\gamma) \over dt} &=&
 {\alem\over 2\pi} e_q^2 \hat P_{q\gamma}(n)
\nonumber\\
&+&
{\alpha\over 2\pi} P_{qq}(n) f^\gamma_{q/\bar q}(n,t,t_\gamma)
 + {\alpha\over 2\pi} P_{qG}(n) f^\gamma_G(n,t,t_\gamma)
,
\\
\label{meg-G}
{df^\gamma_G(n,t,t_\gamma) \over dt} &=& 
{\alpha\over 2\pi} P_{Gq}(n) \sum_{q=1}^{\nf}
\left[ f^\gamma_q(n,t,t_\gamma)+f^\gamma_{\bar q}(n,t,t_\gamma)\right]
\nonumber\\
&+&
{\alpha\over 2\pi} P_{GG}(n) f^\gamma_G(n,t,t_\gamma)
,
\end{eqnarray}
\end{mathletters}
where
\begin{equation}
\hat P_{q\gamma}(x) = 3 [x^2 + (1-x)^2]
\end{equation}
is the photon-quark splitting function and $P_{ik}$ are
the QCD (Altarelli--Parisi) splitting functions.
In these equations the photon virtuality ($t_\gamma$) is fixed
and can be thought of as an ``external'' parameter describing the
state, QCD content of which depends on $t$ and $n$.
The inhomogenous term in \Eq{meg-q} makes the difference
with QCD equations for hadrons.

The standard method of solving the evolution equations is to
decompose first the structure functions into singlet and 
non-singlet components.
To simplify the discussion
we will present formulae for the non-singlet part only.
Defining the non-singlet part of a structure function as
\begin{equation}
\bar f_{q/\bar q} = f_{q/\bar q}
 -{1\over 2\nf} \sum_{q'}^{\nf} (f_q'+f_{\bar q'})
\end{equation}
we obtain
\begin{equation}
{d\fNS \over dt}
=
{\alem\over 2\pi} (e_q^2 - \langle e_q^2\rangle) \hat P_{q\gamma}(n)+
{\alpha\over 2\pi}
 P_{qq}(n) \fNS
,
\label{fqNS}
\end{equation}
where
\begin{equation}
 \langle e_q^2\rangle
= {1\over 2\nf} \sum_{q=1}^{\nf} e_q^2
\end{equation}

In the leading log order of QCD
\begin{equation}
\label{aLL}
{\alpha(t) \over 2 \pi} = {2\over\beta_0 t}
\end{equation}
with $\beta_0 = 11 - 2 n_{\rm f}/3$ for $n_{\rm f}$ flavors,
and
\Eq{fqNS} becomes %now
\begin{equation}
{d\fNS \over dt} = 
\bar d_{q\gamma}(n)
-
{d_{qq}(n)
\over t} \fNS
,
\end{equation}
where
\begin{equation}
\bar d_{q\gamma}(n) = 
{\alem\over 2\pi}
(e_q^2 - \langle e_q^2\rangle)  \hat P_{q\gamma}(n)
\end{equation}
and
\begin{equation}
 d_{qq}(n) = -2{P_{qq}(n)\over \beta_0}
\,.
\end{equation}
The general solution to this differential equation reads
\begin{eqnarray}
\fNS
&=&
{\bar d_{q\gamma}(n)\, t
\over
1 + d_{qq}(n)}
\left[ 1 - \left({t' \over t}\right)^{1+d_{qq}(n)} \right]
\nonumber\\
&&+
\bar f^\gamma_{q/\bar q}(n,t',t_\gamma)
 \, \left({t' \over t}\right)^{d_{qq}(n)}
\label{gamSF}
\end{eqnarray}
As shown for the first time by Witten \cite{resph}, 
the first term is characteristic feature of the photon structure
function and for large $t$ it dominates, resulting in linear growth with
$t$. The second term in this solution depends on the `input' measured at
some $t'$ and is analogous to the leading-log QCD evolution of
hadronic structure functions.
For low $t'$ and $t_\gamma$ the measured structure function
$\bar f^\gamma_{q/\bar q}(n,t',t_\gamma)$ can be identified
with the VDM-like 
$\bar f^{\rm (V)}_{q/\bar q}(n,t',t_\gamma)$ discussed in the
previous section.

\Eq{gamSF} gives no explicit prediction on
the $t_\gamma$ dependence of $\bar f^\gamma_{q/\bar q}(n,t,t_\gamma)$
except for the fact that the asymptotic ($t \gg t'$) solution is
independent of $t_\gamma$
\begin{equation}
\label{eq:gam as}
\fNS \simeq
{\bar d_{q\gamma}(n)
\over
1 + d_{qq}(n)}\, t.
\end{equation}

\begin{figure}
\centerline{\epsfig{file=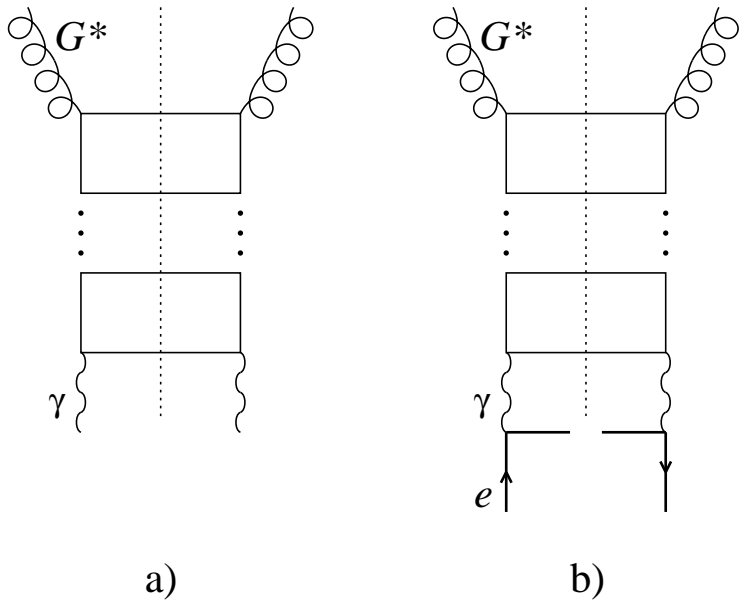,width=10cm}}
\caption{\label{Fgle}
General ladder expansion for the photon (a) 
and the electron structure function (b)}
\end{figure}

Another method of calculating QCD structure functions is the 
ladder expansion. %\cite{DGLAP}. 
The corresponding diagram for the photon structure function
is shown in \Fig{Fgle}a. 
Actual calculations should be performed in the axial gauge
but here we will integrate only over the quark emitted by the photon,
which is as simple as
\begin{equation}
\fNS
=
\bar d_{q\gamma}(n)
\int\limits_{P^2_\gamma}^{Q^2}  {dk^2\over k^2}
f_{qq}(n, Q^2,k^2)
.
\label{gamGdef}
\end{equation}

%FFFFFFFFFFFFFFFFFFFFFFFFFFFFFFFFFFFFFFFFFFFFF

The QCD `structure function of a point-like virtual quark' 
$f_{qq}(n,Q^2,k^2)$ corresponds to the ladder diagram of 
\Fig{Fgle}a without the lowest quark rung.
The latter is explicitly integrated in \Eq{gamGdef} with
the electromagnetic $\gamma$-$q$ coupling contained in 
$\bar d_{q\gamma}(n)$ and $1/k^2$ coming from the quark propagator.
The only difference with pure QCD is that here the coupling 
is electromagnetic and does not depend on $k^2$.
Changing the integration variable to  $\tau = \log (k^2 /\LL)$
and using the QCD formula
\begin{equation}
f_{qq}(n,Q^2,k^2)
=
\left( \tau \over t \right)^{d_{qq}(n)}
\end{equation}
we arrive at \cite{UW}
\begin{eqnarray}
\fNS
&=&
\bar d_{q\gamma}(n)
\int\limits_{t_\gamma}^{t} {d\tau}
\left( \tau \over t \right)^{d_{qq}(n)}
\\
\label{gamGsol}
&=&
{\bar d_{q\gamma}(n)\, t
\over
1 + d_{qq}(n)}
\left[
1 - \left( {t_\gamma \over t} \right)^{1+d_{qq}(n)}
\right].
\end{eqnarray}
Formally this result equals to the solution of master equations
\Eq{gamSF} with $t' = t_\gamma$ and
$\bar f^\gamma_{q/\bar q}(n,t_\gamma,t_\gamma) = 0$.

Let us explain the implicit assumptions 
made in the derivation of \Eq{gamGsol}.
Thanks to the strong ordering of virtualities in the 
ladder expansion
\begin{equation}
P^2_\gamma < k^2 < \dots < Q^2
\end{equation}
we could integrate over the whole range of quark virtualities $k^2$.
Within QCD, however, this can be done only if the photon virtuality
$P^2_\gamma > \LL$.
Moreover we have used the fact that in perturbative calculation
such photon has a point-like coupling to quarks.
In other words the photon of virtuality
$P^2_\gamma > \LL$ has no QCD structure at the scale $Q^2 = P^2_\gamma$.
We see, thus, that \Eq{gamSF} is valid for any $P^2_\gamma$ while \Eq{gamGsol} for
$P^2_\gamma > \LL$ only.
This is exactly the reason for introducing the intermediate 
scale $P_0^2$ in \Eq{qsplot}.

In the following we will assume that $P^2_\gamma$ is large enough
for \Eq{gamGsol} to hold.
With this assumption the integration over $P^2_\gamma$,
 as in \Eq{qsplot1},
 is straightforward and results in the following expression
 for the electron structure function
\begin{eqnarray}
\label{ESF3}
\bar f^e_{q/\bar q}(n, t, t_1, t_0)
&=&
\hat f^e_\gamma(n)
  \int\limits_{t_0}^{t_1}\! dt_\gamma
  \bar f^\gamma_{q/\bar q}(n,t, t_\gamma)
\nonumber\\
&=&
{\hat f^e_\gamma(n) \bar d_{q\gamma}(n)\, t
\over
1 + d_{qq}(n)}
\Bigg\{
t_1-t_0
%....................
\nonumber\\
&&
- {t \over 2 + d_{qq}(n)}
\left[
 \left( {t_1 \over t} \right)^{2+d_{qq}(n)}
-
 \left( {t_0 \over t} \right)^{2+d_{qq}(n)}
\right]
\Bigg\}
.
\end{eqnarray}
This result corresponds to the diagram
depicted in \Fig{Fgle}b, where the 
range of photon virtualities is
controlled by imposing a limit on momentum transfer to the outgoing electron.
\Eq{ESF3} depends on three scales $t > t_1 > t_0$ with $t_0$ kept fixed.
In order to find formulae for large $t$ ($t \gg t_0$)
we have to specify the relation between $t_1$ and $t$.
Let us consider two extreme cases: $t_1 \ll t$ and $t_1 = t$.

\begin{itemize}
\item `Exclusive' case: $t_1 \ll t$
%-----------------------
\begin{equation}
\label{eq:exc}
\bar f^e_{q/\bar q}(n, t, t_1, t_0)
\simeq
{\hat f^e_\gamma(n) \bar d_{q\gamma}(n)
\over 1 + d_{qq}(n) }
\, t (t_1 - t_0).
\end{equation}

\item `Inclusive' case: $t_1 = t$
%-----------------------
\begin{equation}
\label{eq:inc}
\bar f^e_{q/\bar q}(n, t)
\equiv
\bar f^e_{q/\bar q}(n, t, t, t_0)
\simeq
{\hat f^e_\gamma(n) \bar d_{q\gamma}(n)
\over 2 + d_{qq}(n) }
\, t^2
,
\end{equation}
where we have dropped the last two arguments of the `inclusive' structure
function.
\end{itemize}

To obtain the full result for $\bar f^e_{q/\bar q}$ we still have to add
the integral over the photon virtualities below $P_0^2$.
To this end we use \Eq{gamSF} with experimental input
replaced by a VDM-like parametrization at $t'=t_0$:
\begin{eqnarray}
\fNS
&\approx&
{\bar d_{q\gamma}(n)\, t
\over
1 + d_{qq}(n)}
\left[ 1 - \left({t_0 \over t}\right)^{1+d_{qq}(n)} \right]
\nonumber\\
&&+
\bar f^{\rm (V)}_{q/\bar q}(n,t_0,t_\gamma)
 \, \left({t_0 \over t}\right)^{d_{qq}(n)}.
\label{VSF}
\end{eqnarray}
As discussed earlier $\bar f^{\rm (V)}_{q/\bar q}(n,t_0,t_\gamma)$
should decrease with increasing $t_\gamma$ and vanish 
for $t_\gamma \ge t_0$ 
(see {\it e.g.} \cite{GRS} for a phenomenological parametrization).
Thus for $t \gg t_0$ we get the unique prediction independent of
the `input' at low $t_0$
\begin{equation}
\hat f^e_\gamma(n)
  \int\limits_{t_{\rm min}}^{t_0}\! dt_\gamma
  f^\gamma_k(n,t, t_\gamma)
\simeq
{\hat f^e_\gamma(n) \bar d_{q\gamma}(n)\, t
\over
1 + d_{qq}(n)}
(t_0 - t_{\rm min}),
\end{equation}
where $t_{\rm min} = \log (P_{\rm min}^2 /\LL)$.

For large $t$ this low $P_\gamma^2$ contribution grows linearly with $t$
and 
can be neglected in the `inclusive' case.
\Eq{eq:exc} becomes now
\begin{equation}
\label{eq:excf}
\bar f^e_{q/\bar q}(n, t, t_1)
\simeq
{\hat f^e_\gamma(n) \bar d_{q\gamma}(n)
\over 1 + d_{qq}(n) }
\, t (t_1 - t_{\rm min})
\equiv
{\hat f^e_\gamma(n) \bar d_{q\gamma}(n)
\over 1 + d_{qq}(n) }
\, t \, \log{P^2 \over P_{\rm min}^2}
.
\end{equation}
This is exactly the Weizs\"acker-Williams formula \Eq{W-W}
with asymptotic solution \Eq{eq:gam as} 
used for the photon structure function.

Much more interesting is the `inclusive' case \Eq{eq:inc}.
To understand this QCD prediction let us look first at the
photon structure function. There the effect of QCD evolution can be seen by
comparing the full result \Eq{gamGsol} with the Quark Parton Model (QPM) limit.
We reach this limit
by taking $\LL \rightarrow 0$ 
($\alpha(t) \rightarrow 0$) in \Eq{gamGsol},
which results in
\begin{equation}
\label{gQCD}
\fNS \vert_{\rm QPM}
= 
\bar d_{q\gamma}(n)\, (t - t_\gamma)
=
\bar d_{q\gamma}(n)\, \log{Q^2 \over P_\gamma^2}
\end{equation}

Thus we see that for large $t$ the net effect of the QCD evolution 
on the photon structure function
is to change $\bar d_{q\gamma}(n)$ into $\bar d_{q\gamma}(n)
/(1 + d_{qq}(n))$. The dependence on $t$ remains the same
but the structure functions (transformed back to the $x$-space)
have different dependence on $x$. 
As usually the QCD evolution `shifts' the distribution towards lower $x$
values.
Analogously, for the electron structure function the QPM limit of \Eq{ESF3} reads
\begin{eqnarray}
\label{eQPM}
\bar f^e_{q/\bar q}(n, t, t_1, t_0)\vert_{\rm QPM}
&\simeq&
{\hat f^e_\gamma(n) \bar d_{q\gamma}(n)
\over 2}
(t_1 - t_0) (2 t - t_1 -t_0)
\nonumber\\
&=&
{\hat f^e_\gamma(n) \bar d_{q\gamma}(n)
\over 2}
\log {P^2\over P_0^2}
\left( \log {Q^2\over P_0^2} + \log {Q^2\over P^2} \right)
\end{eqnarray}
The reader can easily check that this result corresponds to the integral
$\int_{t_0}^{t_1}\! dt_\gamma$
of the QPM formula for the photon structure function, \Eq{gQCD}.

Comparing the QPM result with the QCD formulae 
\Eq{eq:excf} and \Eq{eq:inc} we see that the effect
 of QCD evolution is
to multiply the moments of the electron structure function
by $1/(1 + d_{qq}(n))$ in the `exclusive' case
and
by $2/(2 + d_{qq}(n))$ in the `inclusive' case.
After transforming back to the $x$-space,
this means that QPM, `exclusive' and `inclusive'
electron structure functions all have different dependence on $x$.
In particular the `inclusive' solution
$\bar f^e_{q/\bar q}(n, t)$, which corresponds to a standard
structure function, gets modified analogously to the photon
case but by another factor.
In this sense the electron acquires an anomalous component from QCD.

So far we have discussed the non-singlet solution. The singlet case
goes along the same lines but will not be presented here. 
Instead, we construct in the next section the evolution equations which
can be solved in the $x$-space.

% SSSSSSSSSSSSSSSSSSSSSSSSSSSSSSSSSSSSSSSS

\section{Evolution equations}
Let us first show that \Eq{ESF3} is the general solution to the
following master equations
\begin{mathletters}
\label{mel}
\begin{eqnarray}
\label{mel-t}
{\partial\feNS \over \partial t} 
&= &
d_{qe}(n) (t_1 - t_0)
-
{d_{qq}(n)
\over t} \feNS
,
\\
\label{mel-t1}
{\partial\feNS \over \partial t_1} 
&= &
\hat f^e_\gamma(n) \bar f^\gamma_{q/\bar q}(n,t,t_1)
\nonumber\\
&= &
{d_{qe}(n)\, t
\over
1 + d_{qq}(n)}
\left[
1 - \left( {t_1 \over t} \right)^{d_{qq}(n)+1}
\right]
,
\end{eqnarray}
\end{mathletters}
where $d_{qe}(n) \equiv \hat f^e_\gamma(n) \bar d_{q\gamma}(n)$.
Note that the second equation is just the derivative of \Eq{qsplot}
with \Eq{gamGsol} inserted for the photon structure function.
A general solution to \Eq{mel-t} %the first equation
can be written as
\begin{equation}
\feNS
=
C(n,t_1) \, t^{-d_{qq}(n)}
+
{d_{qe}(n)
\over
1 + d_{qq}(n)}\, (t_1 - t_0) t.
\label{eSF0}
\end{equation}
So far $C(n,t_1)$ is an arbitrary function of $t_1$.
Nb. if $t_1$ remains constant when $t \rightarrow \infty$
the second term of \Eq{eSF0} gives the asymptotic solution for the
`exclusive' case.

From the second equation \Eq{mel-t1} we obtain
\begin{equation}
C(n,t_1)
= C^{(0)}(n)
- {d_{qe}(n) \, t_1^{d_{qq}(n)+2}
\over [1 + d_{qq}(n)]\,[2 + d_{qq}(n)]}
\end{equation}
with arbitrary $C^{(0)}(n)$.

Now the general solution to \Eq{mel} reads
\begin{eqnarray}
\feNS
&=&
C^{(0)}(n) t^{-d_{qq}(n)} 
\nonumber\\
&&+ {d_{qe}(n) \over 1 + d_{qq}(n)}
\left[ 
(t_1 - t_0) t 
- 
{ t_1^{2+d_{qq}(n)}  t^{-d_{qq}(n)} 
\over 2 + d_{qq}(n)}
\right].
\label{eSF1}
\end{eqnarray}
Upon imposing the boundary condition
$\bar f^e_{q/\bar q}(n, t, t_0, t_0)=0$
we recover the formula \Eq{ESF3} derived in the previous section.

The complete set of master equations in the $x$-space
analogous to \Eq{mel}
can be obtained by
changing the products of moments into convolutions
and observing that there is no inhomogenous term for the
gluonic component --- {\it c.f.}{} \Eq{meg}.
Suppressing explicit $t_0$ dependence in the function arguments 
we have
\begin{mathletters}
\begin{eqnarray}
{\partial f^e_{q/\bar q}(t,t_1) \over \partial t}
&=&
{\alem\over 2\pi} \hat P_{qe} [t_1 - t_0]
\nonumber\\
&&+
{\alpha\over 2\pi} %\left[
 P_{qq} \otimes f^e_{q/\bar q}(t,t_1) + 
{\alpha\over 2\pi} P_{qG} \otimes f^e_G(t,t_1)
,
\\
{\partial f^e_{q/\bar q}(t,t_1) \over \partial t_1}
&=&
\hat f^e_\gamma \otimes f^\gamma_{q/\bar q}(t,t_1),
\\
{\partial f^e_G(t,t_1) \over \partial t} &=& 
{\alpha\over 2\pi} %\left[
P_{Gq} \otimes \sum_{q=1}^{\nf}
\left[ f^e_q(t,t_1) + f^e_{\bar q}(t,t_1) \right] 
\nonumber\\
&&+
{\alpha\over 2\pi}  P_{GG} \otimes f^e_G(t,t_1)
,
\\
{\partial f^e_G(t,t_1) \over \partial t_1}
&=&
\hat f^e_\gamma \otimes f^\gamma_G(t,t_1),
\end{eqnarray}
\end{mathletters}
where
$\hat P_{qe} = \hat P_{q \gamma} \otimes \hat f^e_\gamma$.

These equations, as they stand, are not very useful because they
contain both electron and photon structure functions.
We will present now how a closed set of equations for 
the electron structure function is formed in the large $t$ region.
As already discussed previously, the large $t$ limit
depends on additional assumptions on $t_1$ vs. $t$ dependence.
Quite generally one can take $t_1$ to be some function of $t$
defining a `path' in the $t$-$t_1$ plane along which the large $t$ limit
is approached. Denoting $t_1 = g(t)$ we have
\begin{eqnarray}
{df^e_k(t,t_1) \over dt}
&=&
{\partial f^e_k(t,t_1) \over \partial t}
+ g'(t) {\partial f^e_k(t,t_1) \over \partial t_1}
\nonumber\\
&=&
{\partial f^e_k(t,t_1) \over \partial t}
+ g'(t)
\hat f^e_\gamma \otimes f^\gamma_k(t,t_1).
\label{eevol}
\end{eqnarray}

A simple choice for $g(t)$ is
\begin{equation}
g(t) = (1-a) \hat t_1 + at
\end{equation}
where $\hat t_1$ is constant and $0 \le a \le 1$.
The two extreme cases considered in the previous section
correspond to following choices for the parameter $a$
\begin{itemize}
\item `Exclusive' case\\
$a=0$, {\it i.e.} $g(t) = \hat t_1$.
As $g'(t) = 0$ the second term in \Eq{eevol} vanishes.
The solution grows linearly with $t$ ({\it c.f.} \Eq{ESF0}).
The resulting
master equations have a constant ($t$-independent) inhomogenous term
\begin{mathletters}
\label{master}
\begin{eqnarray}
{df^e_{q/\bar q}(t,\hat t_1) \over dt}
&=&
{\alem\over 2\pi} \hat P_{qe} [\hat t_1 - t_0]
\nonumber\\
&&+
{\alpha\over 2\pi} %\left[
 P_{qq} \otimes f^e_{q/\bar q}(t,\hat t_1) + 
{\alpha\over 2\pi} P_{qG} \otimes f^e_G(t,\hat t_1)
,
\\
{df^e_G(t,\hat t_1) \over dt} &=& 
{\alpha\over 2\pi}
P_{Gq} \otimes \sum_{q=1}^{\nf}
\left[ f^e_q(t,\hat t_1) + f^e_{\bar q}(t,\hat t_1) \right] 
\nonumber\\
&&+
{\alpha\over 2\pi}  P_{GG} \otimes f^e_G(t,\hat t_1)
\end{eqnarray}
\end{mathletters}
\item `Inclusive' case.\\
$a=1$, {\it i.e.} $g(t)=t$ 
and $g'(t) = 1$.
Now the virtual photon structure function $f^\gamma_k(t,t)$
vanishes because both arguments are equal. 
We arrive at the equations which are formally the same as in the
`exclusive' case but with $\hat t_1$ set to $t$ and $t_{\rm min}$ neglected.
The master equations have now the inhomogenous term proportional to 
$t$, which results in a different $x$-dependence of the solutions.
\end{itemize}

According to \Eq{eq:exc} and \Eq{eq:inc}
the asymptotic solutions to these equations have the form
\begin{equation}
\label{aseq}
f^e_{k}(z,t) = 
\left( {\alem\over 2\pi} \right)^2
\tilde f^e_{k}(z)\, t t_1
\end{equation}
with $t_1$ constant and $t_1=t$ for the `exclusive' and `inclusive' 
case, respectively. Here we have taken $t_0 = 0$ ($P_0^2 = \LL$).
Substituting the leading-log formula \Eq{aLL} for $\alpha(t)$ we 
end up with $t$-independent integral equations
\begin{mathletters}
\begin{eqnarray}
p \tilde f^e_{q/\bar q}(z)
&=&
\hat P_{qe}(z)
+
{2 \over \beta_0} 
\int\limits_z^1\!{dx\over x}
\left[
 P_{qq}(x) \tilde f^e_{q/\bar q}\!\left({z\over x}\right) + 
P_{qG}(x) \tilde f^e_G\!\left({z\over x}\right)
\right]\!,
\\
p \tilde f^e_G(z) &=& 
{2 \over \beta_0} 
\int\limits_z^1 \!{dx\over x}
\sum_{k=q,\bar q,G}
P_{kq}(x) 
  \tilde f^e_k\left({z\over x}\right),
\end{eqnarray}
\end{mathletters}
where $p=1$ or $p=2$ for the `exclusive' or `inclusive' 
case, respectively.

%FFFFFFFFFFFFFFFFFFFFFFFFFFFFFFFFFFFFFFFFFFFFF
\begin{figure}
\centerline{\epsfig{file=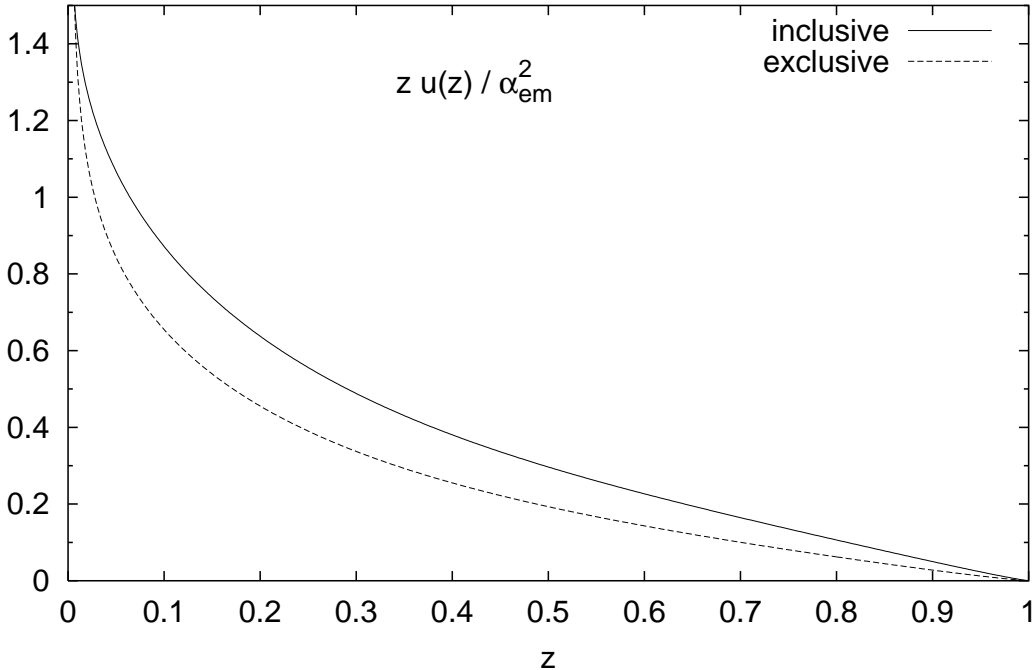,width=14cm}}
\caption{\label{Fu}
Asymptotic $u$-quark `exclusive' and `inclusive' 
distributions $zf^e_u(z,Q^2)/\alem^2$ for $Q^2 = 100{\rm GeV}^2$
and $P^2 = 1{\rm GeV}^2$ for `exclusive' case. 
Calculations have been done for 5 flavors, 
$\Lambda_{\rm QCD} = 0.2{\rm GeV}$ and $P_0^2 = \LL$.}
\end{figure}

We solve these equations numerically using the method described
in \cite{wjWBSF}. In \Fig{Fu} we show the 
`exclusive' and `inclusive' solutions for the distribution
of $u$ quarks in the electron at $Q^2 = 100{\rm GeV}^2$.
There is still the contribution from the photons of low virtuality
($P_\gamma^2 < \LL$)
which should be added, as it has been done in \Eq{eq:excf}.
At asymptotically large $t = \log(Q^2/\LL)$ it modifies the `exclusive' solution
only, by changing the $t_1 = \log(P^2/\LL)$ factor in
\Eq{aseq} into $\log(P^2/P_{\rm min}^2)$.
At finite $Q^2$, however, this contribution is non-negligible for
`inclusive' case and should also be added.
As can be seen from \Eq{ESF3} taking the intermediate scale 
$P_0^2 = \LL$ ($t_0=0$) sets the proper normalization 
for this correction%
\footnote{At $Q^2$ value considered here the low $P_\gamma^2$ contribution
is of the same order of magnitude as the QCD results in \Fig{Fu}.}
at finite $t$.
Thus the both curves in \Fig{Fu} get shifted by the same amount.

\section{Summary}
In this paper we have analyzed the parton content of the electron within
perturbative QCD. We have shown that electron acquires an anomalous
component from QCD, analogously to photon.
We have constructed the evolution equations for the `exclusive' and
`inclusive' electron structure function. 
These two cases correspond to `anti-tagging' and `no-tagging'
experimental conditions, respectively.
The evolution equations can be solved numerically in the $x$-space in
the asymptotic $Q^2$ region. 
As an example we have shown the $u$ quark distribution
inside the electron.

The results presented here are leading-log QCD solutions valid at
asymptotically large $Q^2$.
At finite $Q^2$ the next-to-leading corrections, as well as
non-perturbative contributions including hadronic component of the real photon
will modify the results.
Despite these inaccuracies it would be interesting 
to compare these predictions with experiment.
On one hand,
the data for the electron structure function should be much more precise 
than the ones used for the photon structure function.
On the other hand,
the improvements on the theoretical side can be done in a similar way
as for the photon structure function ---
higher order 
perturbative %and non-perturbative
QCD effects, as well as phenomenological parametrizations can be plugged
into the evolution equations \Eq{master}.

\bigskip

The author would like to thank Jerzy Szwed for numerous discussions 
and for critical reading of the manuscript.
The hospitality of DESY Theory group, where part of this work was done,
is also acknowledged.

%-------------------------------------------------

%===================================================
\end{document}